\documentclass[]{article}

\usepackage{graphicx}
\usepackage{dcolumn}
\usepackage{bm}
\usepackage[utf8]{inputenc}
\usepackage[T1]{fontenc}
\usepackage{mathptmx}
\usepackage{amsfonts}
\usepackage{bbm}
\usepackage[bottom]{footmisc}
\usepackage[utf8]{inputenc}
\usepackage{amsmath}
\usepackage{amssymb}
\usepackage{amsfonts}
\usepackage{amsthm}

\fontsize{11}{16}\selectfont

\begin{document}

\title{A new foundation of quantum decision theory}

\author{Inge S. Helland, Department of Mathematics. University of Oslo\\ P.O. Box 1053, N-0316 Oslo, Norway\\ ingeh@math.uio.no\\ ORCID: 0000-0002-7136-873X }

\date{}

\maketitle

\begin{abstract}
Quantum decision theory is introduced here, and new basis for this theory is proposed. It is first based upon the author's general arguments for the Hilbert space formalism in quantum theory, next on arguments for the Born rule, that is, the basis for calculating quantum probabilities. A basic notion behind the quantum theory foundation is that of theoretical variables, that are divided into accessible and inaccessible ones. This is here specialized to decision variables. It is assumed that all accessible variables can be seen as functions of a specific inaccessible variable. Another assumption is that there exist two different maximal accessible theoretical variables in the given situation. Two basic assumptions behind the Born rule are 1) the likelihood principle, 2) the actor in question has motivations that can be modeled by a hypothetical perfectly rational higher being. The theory is illustrated by a medical example. Finally, a brief discussion of decision processes is given.
\end{abstract}


\section{Introduction}

The notion of `decision' is fundamental to all sciences. The notion is analysed in psychological articles and books, but it also lies in the foundation of much economic theory and statistical theory, where optimal decisions often are coupled to expected utility. Of course, in addition the notion has its daily life implications.

Decisions can be made on the basis of knowledge, on the basis of beliefs, or both. They are always made in a concrete context. Single persons can make decisions, and joint decisions can be made by a group of communicating persons.

Political decisions are important, and may have large consequences. In an autocratic country political decisions are made by sovereign leaders, having gained their positions through power exchanges. In democratic countries, at least in principle people may control their leaders' decisions.

In this article I will concentrate on decisions made by single persons. From a psychological point of view, such decisions have been thoroughly reviewed by Newell et al. (2015). The book emphasizes the relationship between learning and decisions, arguing that the best way to understand how and why decisions are made is in the context of learning and knowledge acquisition which precedes them, and the feedback which follows. Some of the discussions in op. cit. are also relevant for the developments in the present article.

Knowledge acquisition is fundamental to the process of making judgements, which precedes any mature decision. According to Newell et al. (2015), knowledge acquisition can be divided into 3 steps: 1) Discovering information; 2) Receiving and searching through information; 3) Combining information.The first step is particularly important, and is modelled through Brunswik's lens model: Variables in the real world go through `lenses of cues' before they are perceived by our minds. In a simplified way we can say that we discover information by asking questions to nature, and receive answers. The different questions may be coupled.

A person always has his history, and through his history he learns how to handle decisions. Hence learning is important. A person's learning history starts when he or she is a child, when his or her ideals usually are the parents. Later, he or she is influenced by other people, but we can never disregard such influences as a background for a person's decisions, even in cases where the person claims that his aim is just to optimize his behaviour according to some criterion.

In standard economic theory and statistical theory, the problem of optimizing behaviour is reduced to optimizing expected reward or utility. I will claim that this does not quite correspond to the way most people act during a decision process. This statement can be supported by a number of empirical investigations; see for instance the review articles mentioned below in connection to quantum decision theory. The argument will be further elucidated later in this article.

Admittedly, the decision foundation based upon optimizing expected utility has had and has still a strong theoretical support. There are several axiom systems that can motivate such a basis, and the mathematics behind such axioms can be traced back to Bernoulli and Pascal. Newell et al. (2015) discuss one such axiom, the sure thing principle: If a person prefers option A to option B under the condition X, and he also prefers A to B under the opposite condition not-X, then he always prefers A to B.

One difficulty with the sure thing principle, is that one can find examples where rational people do not seem to follow this principle. Two such examples are given by Allais (1953) and Ellsberg (1961).

A rather primitive form of learning is reinforcement learning, learning governed by rewards or punishments.
In Friston et al. (2009) the notion of active inference was introduced as an alternative to reinforcement learning in connection to decision making. This may be important, since the way many people learn in connection to decisions obviously is more active than what can be described as reinforcement learning. In op cit. a theory of active inference is formalized in terms of a free-energy formulation of perception, quantizised in terms of entropy or average surprise associated with a probability distribution on an agent's state space and his environment.

In Kahneman et al. (2021) decisions are discussed in terms of their noise and skewness. Both are problematic sides of human decisions. Skewness denotes systematic errors made in relation to some defined goal. Noise denotes variability: Imagine two medical doctors in the same city giving different diagnoses to identical patients. Imagine two judges assigning different punishments for the same crimes. Or imagine the same medical doctor or the same judge getting at different results depending upon whether it is before or after lunch, or Monday instead of Friday. According to op. cit. our decisions, also decisions made by socalled experts, have larger random variation than we usually think of. Means for reducing noise are discussed in the book.

Theories of decisions can take many points of departure. In this article we will describe a new foundation of a comparatively recent theory: Quantum decision theory (QDT). This foundation is completely based upon a new foundation of quantum mechanics, discussed in the book Helland (2021) and in some newer articles, Helland (2022a,b,c, 2023a,b,c, 2024a,b,c, 2025a,b, 2026).

Pionering articles on quantum decision theory are Aerts (2009) and Yukalov and Sornette (2014); see also the book by Busemeyer and Bruza (2012). More recent review articles are Ashtiani and Azgomi (2015) and Pothos an Busemeyer (2022). A motivation and more references can also be found in the article by Wang et al. (2013). My approach will be independent of this literature; a completely new entrance to the theory will be aimed at.

The plan of this article is as follows: I first review briefly the new approach to the foundation of quantum theory, and then I use this to derive and motivate a new foundation of quantum decision theory, taking as a basis a simple model of a person's decisions. First, the Hilbert space formalism of decision variables is derived. Then the Born formula, the basis for computing quantum probabilities, is argued for under weak assumptions, and this is specialized to a decision setting. An example of applications is then given, and used to illustrate some of the peculiarities of quantum probabilities. Then  my foundation of QDT is further discussed from several points of view before I give some concluding remarks.

\section{A new foundation of quantum decision theory}

\subsection{On the foundation of quantum theory in general}

A completely new approach towards quantum foundations is proposed in Helland (2021, 2022a,b,c 2023a,b,c,2024a,b,c, 2025a). The basis of this approach is an observer who is in some physical situation. In this situation there are theoretical variables, and certain of these variables, say $\theta, \lambda, \eta,...$ are related to the observer $C$. Some of the variables are \emph{accessible} to him, meaning roughly that it is, in some future, in principle possible to obtain as accurate values as he wishes on the relevant variable. Others are \emph{inaccessible}. Examples of the latter are the vector (position, momentum) of a particle at some time, or the full spin vector of a spin particle, an imagined vector whose discretized projection in the direction $a$ is the spin component in that direction. The terms `accessible' and `inaccessible' are primitive notions of the theory.

The main assumption of my theory is then as follows: \emph{Related to an observer $C$ there is an inaccessible variable $\phi$ such that all accessible variables can be seen as functions of $\phi$}. In the two examples above one can take $\phi =$ (position, momentum) and $\phi =$ full spin vector. In the last example, say in the spin 1/2 case, one can model the discrete spin component $\theta^a$ in direction $a$ as $\mathrm{sign}(\mathrm{cos}(a, \phi))$. Giving $\phi$ a reasonable distribution here, results in a correct distribution of each $\theta^a$. A symmetry assumption, needed for continuous variables, is that there exists a group $K$ acting on the range of $\phi$.

Here these variables are theoretical  variables coupled to some physical situation. If necessary, an observed variable can be modeled as a theoretical variable plus some random error. Following Zwirn (2016, 2020), every description of reality must be seen from the point of view of some observer. Hence we can assume that the variables also exist relative to $C$.

But observers may communicate. The mathematical model developed in the articles above is equally valid relative to a group of people that can communicate about the physics and about the various theoretical variables. This gives a new version of the theory, a version where all theoretical variables are defined jointly for such a group. The only difference here is that, for the variables to function during the communication, they must always be possible to define them in words.

In the above two examples there are also maximal accessible variables: In the first example either position or momentum, in the second example the spin component $\theta^a$ in some direction $a$. From a mathematical point of view, an accessible variable $\theta$ is called maximal if there is no other accessible variable $\lambda$ such that $\theta = f(\lambda)$ for some non-invertible function $f$. In general, the term `maximal' will then be seen to be maximal with respect to the partial ordering of variables given by $\alpha\le\beta$ iff $\alpha=f(\beta)$ for some function $f$.

Two different acccessible variables $\theta$ and $\eta$ are said to be related if there is a transformation $k$ in $\phi$-space and a function $f$ such that $\theta = f(\phi)$ and $\eta = f(k\phi)$. In Helland (2024a) it is shown that if both $\theta$ and $\eta$ are maximal and take $r$ values, then $\phi$ can be constructed in a natural way such that there is an accessible variable $\lambda$ that is in one-to-one correspondence with $\theta$, and such that $\lambda$ and $\eta$ are related according to this definition. 

 In this sense, one can always think of some relation between $\theta$ and $\eta$ even when they are different maximal accessible variables. Variables that are maximal, different, have similar range spaces, and not in one-to-one correspondence, are called - following Niels Bohr - for complementary.

In many cases, in particular in the two examples above, $\lambda$ and $\theta$ are identical, so that $\theta$ and $\eta$ are directly related. Two spin components $\theta^a$ and $\theta^b$ are related, and position and momentum are related theoretical variables. In the first case, $\phi$-space can be taken as the plane spanned by the directions $a$ and $b$, and $k$ can be taken as a $180^o$ rotation around the midline between $a$ and $b$. In the last case, $k$ is constructed by a Fourier transform.

In the present paper, I will concentrate on accessible variables that take a finite number of values, say $u_1,u_2,...,u_r$. Then it can be shown that pairs of maximal accessible variables are either related, are in one-to-one correspondence or have a combination of these two properties. The following Theorem is proved in Helland (2024a): 
\bigskip

\textbf{Theorem 1} \textit{Assume that there exist two different maximal accessible variables $\theta$ and $\eta$, each taking $r$ values, and not in one-to-one correpondence. Then, there exists an $r$-dimensional Hilbert space $\mathcal{H}$ describing the situation, and every accessible variable in this situation will have an associated self-adjoint operator in $\mathcal{H}$.}
\bigskip

This theorem is the first step in a new proposed foundation of quantum theory. 

The second step is to show the following: If $k$ is the transformation connecting two related variables $\theta$ and $\eta$, then there is a unitary matrix $W(k)$ such that $A^\eta = W(k)^{-1} A^\theta W(k)$. 

Given these results, a rich theory follows. The set of eigenvalues of the operator $A^\theta$ is identical to the set of possible values of $\theta$. The variable $\theta$ is maximal if and only if all eigenvalues of the corresponding operator are simple. In general, the eigenspaces of $A^\theta$ are in one-to-one correspondence with questions `What is $\theta$'/ `What will $\theta$ be if we measure it?' together with sharp answers $\theta=u$, where $u$ is the relevant  eigenvalue of $A^\theta$.

If $\theta$ is maximal as an accessible variable, the eigenvectors of $A^\theta$ have a similar interpretation. In my opinion, such eigenvectors, where $\theta$ is some meaningful variable, should be tried to be taken as the only possible state vectors. These have straightforward interpretations, and from this version of the theory, also a number of so-called `quantum paradoxes' can be illuminated, see Helland (2023a,c).

This not only points at a new foundation of quantum theory, and it also suggests a general epistemic interpretation of the theory: Quantum theory is not directly a theory about the world, but a theory about an actor's knowledge of the world. Versions of such an interpretation already exist, and they are among the very many suggested interpretations of quantum mechanics.

Below, I will need some mathematical notations from standard finite-dimensional quantum mechanics, notations mainly due to Dirac:

A Hilbert space $\mathcal{H}$ can be taken as an $r$-dimensional complex vector space. Operators on this space are then $r\times r$ matrices. An operator $A$ is said to be self-adjoint if $A^\dagger = A$, where $A^\dagger$ is formed from $A$ by transposing it and taking complex conjugates of all elements. Self-adjoint operators have real-valued eigenvalues. An operator $U$ is unitary if $U^{-1} = U^\dagger$.

A ket vector $|\psi\rangle$, a state vector, can be seen as a unit column vector in $\mathcal{H}$. I will concentrate on state vectors of the form $|\psi\rangle = |b;j\rangle$, an eigenvector of some operator $A^b$, in my language, equivalent to a particular answer to a question of the form `What is $\theta^b$?', where $\theta^b$ is a maximal accessible theoretical variable.

To any ket vektor $|a;k\rangle$ there corresponds a bra vector $\langle a;k|$, the corresponding complex conjugate row vector. Scalar products between two state vectors can then be written $\langle a;k |b;j\rangle$. I will also need scalar products of the form $\langle a;k |A|a;k\rangle$, where $A$ is an operator.

The projection operator $\Pi_j^b =|b;j\rangle\langle b;j|$ projects onto the state vector $|b;j\rangle$. In general, a projection operator $\Pi_B$ projects onto the line, plane or hyperplane $B$.

\subsection{A simple model of a person's decisions}

Consider a person $C$ in some decision situation. Say that he or she has the choice between a finite set of actions $a_1,..., a_r$. Relative to this situation we can define a finite-valued decision variable $\theta$, taking the different values $1,...,r$, such that $\theta=i$ corresponds to the action $a_i$ $(i=1,..., r)$. If $C$ really is able to make a decision here and carry out the actions, we say that $\theta$ is an accessible variable. In analogy with the situation in the previous subsection, the variable $\theta$ is in relation to a person $C$, in fact, here $\theta$ belongs to the mind of $C$.

I will regard accessible decision variables as special accessible variables. They may be functions of other theoretical variables that are not necessarily decision variables. But note that any decision variable, $\theta$ say, corresponding to possible actions $a_1, ... ,a_r$, may be seen as a function of another decision variable: Just expand the set of possible actions. And there are other decision variables that are functions of $\theta$: Collect the actions $\{a_i\}$ into subsets of these actions.

In the general theory of Subsection 2.1, every accessible variable $\theta$ is assumed to be a function of a maximal accessible variable $\eta$. When $\theta$ is a decision variable, we can, without loss of generalty assume that this $\eta$ also is a decision variable.
\bigskip

\textbf{Lemma 1} \textit{Let $\theta$ be an accessible decision variable. Then there exists a maximal accessible decision variable $\eta$ and a decision variable $\zeta$ which is a one-to-one function of $\theta$ such that $\zeta\le \eta$ in the partial ordering defined by functional dependence.}
\bigskip

\underline{Proof.} By assumption, there exists a maximal variable $\xi$ such that $\theta\le \xi$, that is, $\theta$ is a function of $\xi$. The crucial point is that $\xi$ always can be seen as a one-to-one function of a decision variable $\eta$: Without loss of generality, let the different values of $\xi$ be $1, ... ,s$ for some $s\ge r$, the different actions in $\theta$. Define a decision process for $C$ by taking some arbitrary action $b_j$ when $\xi = j$, $(j= 1, ... ,s)$. This defines the decision variable $\eta$. Since we must have $s\ge r$, one can collect the actions $b_j$ in subsets that are in one-to-one correspondence with $\{a_i ; i=1, ... ,r\}$. This defines the decision variable $\zeta$.

\qed
\bigskip

We all go through life making decision after decision. Some of these are simple, but some can be really demanding. For the different decisions that $C$ is about to make at some time $t$, there correspond decision variables $\theta, \eta, \lambda,...$. Some of these may be accessible to $C$, but some may be inaccessible: In the given situation, $C$ is simply not able to make up his mind. Consider a fixed time $t$ and let $C$ be in some concrete situation at time $t$. Assume that he at time $t$ is faced with two different maximal decision problems, both corresponding to $r$ different actions.

This must be made precise. A decision problem is said to be maximal if $C$ is just able to make his mind with respect to this decision; if the problem is made slightly more complicated, he is not able to take a decision. Let two different maximal  decision variables be $\theta$ and $\eta$, where $\theta=i$ corresponds to the action $a_i$ $(i=1,...,r)$, and $\eta=j$ corresponds to the action $b_j$ $(j=1,...,r)$. Then, by the abstract mathematical version of the theory of the previous Subsection, we can model the situation by using quantum mechanics.

Note that both $\theta$ and $\eta$ may be vector variables. Say that $\theta=(\theta_1,..., \theta_n)$, where each $\theta_j$ is a simpler decision variable. Then this corresponds to a situation where $C$, in addition to the difficult decision given by $\eta$, is faced with $n$ more simple decisions. Such a situation is not uncommon. In each situation where we shall make a difficult decision, we will be in a context where also a number of trivial decisions may have to be made just in order to survive and to function well in the given context. For many people, these trivial decision occupy a large portion of their mind, such a large part that the vector decision variable $\theta$ also must be considered to be maximal. The assumption that both $\theta$ and $\eta$ take the same number of values $r$, can be satisfied by artificially adding some actions to one of the decision problems.

The simple model above does not cover all situations. Sometimes we have a choice between an infinite number of possibilities, and sometimes the outer context changes during the decision process. Nevertheless, the simple model is a good starting point.

Another simple model is the following: Let $\eta$ be a maximal accessible variable of some kind, describing the relevant state of the observer $C$. For simplicity, assume that $\eta$ takes a finite number $s$ of values. Let $\theta$ be a maximal accessible decision variable taking $r$ values, which can be extended to $s$ by adding some impossible actions. Theorem 1 is equally valid in this situation.

In this article, I will mainly focus on maximal decision problems. A thorough discussion including decisions that are not maximal, is given in Mogiliansky et al. (2009). By Lemma 1, each decision variable may be seen as a function of a maximal accessible decision variable.

 I will maintain that the term maximal has a precise definition, as given in Subsection 2.1. To repeat , let first a partial ordering among theoretical variables be defined by $\alpha\le\beta$ if $\alpha = f(\beta)$ for some function $f$. An accessible variable $\theta$ is maximal if it is maximal among the accessible variables with respect to this partial ordering.

In a setting with decision variables, such a variable can be seen to be maximal if the corresponding actor(s) just is/are able to make the decision. If one adds one extra possible action, $C$ is then unable to make the decision.

\subsection{The Hilbert space formalism for decisions}

It is well known that our minds may be limited, for instance when faced with difficult decisions. I will first mention a side result in this direction from the present development.

 In Helland (2022c), Theorem 2 says essentially: Imagine a person $C$ which in some context has two related maximal accessible variables $\theta$ and $\eta$ in his mind. Impose a specific symmetry assumption. Then $C$ cannot simultaneous have in mind any other maximal accessible varable which is related to $\theta$, but not related to $\eta$. It was claimed in Helland (2022c) that the violation of a famous inequality by practical Bell experiments, can be understood on the basis of this theorem. See also Helland (2023b), where a corresponding theorem is formulated without any symmetry assumption in the case where the maximal accessible variables take a finite number of values.
 
Note that this result has the qualification `at the same time', and indicate a specific restriction to the two maximal variables. But the human mind is very flexible. Taking time into account, we can think of very many variables, even ones that are not related.

For the present article, however, the direct results from Helland (2022b, 2024a) are equally important. Consider again a decision situation, and assume the simple model of the previous Subsection. In particular let $C$ at the same time be confronted with at least two different maximal related decision processes. Then the following hold:

- Each decision variable $\eta$ is associated with a self-adjoint operator $A^\eta$, whose eigenvalues are the  possible values of $\eta$.

- The decision process is maximal if and only if each eigenvalue of the corresponding operator is single.

- In the maximal case, the eigenvectors of the operator can be given interpretations: They are coupled to one particular decision process and a specific choice in this decision process: In concrete terms, the eigenvectors $v$ are in one-to-one correspondence with 1) some maximal accessible decision variable $\eta$, and 2) a specific value $u$ of $\eta$. In other words, the possible eigenvectors are in one-to-one correspondence with 1) the question `Which decision process?' and 2) `Which action did this decision process lead to?'.

- In the general case, the eigenspaces of the operator have a similar interpretation.

This can be taken as a starting point of quantum decision theory, but to develop this theory further, we need to be able to calculate probabilities for the various decisions.

\subsection{The Born rule; general arguments}

Born's formula is the basis for all probability calculations in quantum mechanics. In textbooks it is usually stated as a separate axiom, but it has also been argued for by
using various sets of assumptions; see Campanella et al. (2020) for some references. In fact, the first argument for the Born formula, assuming that there is an affine mapping from set of density functions to the corresponding probability functions, is due to von Neumann (1927). Here we will try to use assumptions that are as weak as possible, and assumptions that can be related to notions both from statistical theory and quantum theory. We start with assuming the likelihood principle: The experimental evidence from any experiment must always be based upon the likelihood $l$, which is probability density or the point probability of observations, seen as a function of the full parameter.

The likelihood concept is then generalized to a quantum setting; After an experiment is done, and given some context $\tau$, all evidence on the maximal parameter $\theta^b$ is contained in the likelihood
$p(z^b|\tau ,\theta^b )$, where $z^b$ is the data relevant for inference on $\theta^b$, also assumed discrete. This is summarized in the
\emph{likelihood effect}:
\begin{equation}
F^b(\mathbf{u}^b ; z^b ,\tau)=\sum_j p(z^b |\tau ,\theta^b =u^b_j ) |b;j\rangle\langle b;j|,
\label{xx}
\end{equation}
where the pure state $|b;j\rangle$ corresponds to the event $\theta^b = u^b_j$, and where $\mathbf{u}^b = (u^b_1,...,u^b_r)$ is a vector of actual or potential values of the parameter $\theta^b$.
\bigskip

The interpretation of the likelihood effect $F^b(\mathbf{u}^b; z^b ,\tau)$ can be formulated as follows:
(1) We have posed some inference question on the accessible conceptual variable
 $\theta^b$. (2) We have specified the relevant likelihood for the data. The question itself and the likelihood for all possible answers of the question, perhaps formulated in terms of state
vectors, can be recovered from the likelihood effect.

The likelihood effect is closely connected to the concept of an operator-valued measure; see a discussion in Helland (2021). Since the focused question here assumed discrete data, each likelihood is in the range $0\le p \le 1$. In the quantum mechanical literature, an effect is any operator
with eigenvalues in the range $[0,1]$.

I will base the discussion upon the following assumption and theorem from Helland (2021), where a further discussion is given.
\bigskip

\textbf{Assumption 1}
\textit{Consider in the context $\tau$ an epistemic setting where the the likelihood principle from statistics is satisfied, and the whole
situation is acted upon by an actor $C$ whose decisions can be modelled to be influenced by a superior being $D$. Assume that $D$'s relevant probabilities for the situation are given by the numbers $q$, and that $D$ can be seen to be perfectly rational in agreement with the Dutch Book Principle.}
\bigskip

The Dutch Book Principle says as follows: No choice of payoffs in a series of bets shall lead to a sure loss for the bettor.

A situation where Assumption 1  holds will be called a \textit{rational epistemic setting}. It will be seen to imply essential aspects of quantum mechanics. Below we will discuss whether or not it also can be coupled to certain macroscopic situations, in particular decision situations.

In Helland (2021), a generalized likelihood principle is proved from the ordinary likelihood principle: Given some experiement, or more generally, some context $\tau$ connected to an experiment, any experimental evidence will under weak assumptions be a function of the likelihood effect $F$. In particular, the relevant probabilities $q$ are functions of $F$: $q(F|\tau)$.

The following Theorem and its Corollary are proved in Helland (2021):
\bigskip

\textbf{Theorem 2} \textit{Assume a rational epistemic setting, and assume a fixed context $\tau$. Let $F_{1}$ and $F_{2}$ be two likelihood effects in this setting,
and assume that $F_{1}+F_{2}$ also
is an effect. Then the experimental evidences, taken as the epistemic probabilities related to the data of the performed experiments, satisfy}
\[q(F_{1}+F_{2}|\tau)=q(F_{1}|\tau)+q(F_{2}|\tau).\]

\bigskip

\textbf{Corollary 1} \textit{Assume a rational epistemic setting in the context $\tau$. Let $F_{1}$, $F_{2}$, \ldots be likelihood effects in this setting, and assume that $F_{1}+F_{2}+\ldots$ also
is an effect. Then}
\[q(F_{1}+F_{2}+\ldots|\tau) = q(F_{1}|\tau)+q(F_{2}|\tau)+\ldots.\]

 The further derivations rely on a very elegant recent theorem by Busch (2003): 
Let in general $\mathcal{H}$ be any separable Hilbert space. Recall that an
effect $F$ is any operator on the Hilbert space with eigenvalues in the range $[0,1]$. A generalized probability measure $\mu$ is a function on the effects with the properties
\[
\begin{array}{l}
(1)\ 0\le \mu(F)\le 1\ \mathrm{for\ all}\ F,\\
(2)\ \mu(I)=1,\\
(3)\ \mu(F_{1}+F_{2}+\ldots)=\mu(F_{1})+\mu(F_{2})+\ldots\ \mathrm{whenever}\ F_{1}+F_{2}+\ldots\le I.
\end{array}
\]
\bigskip

\textbf{Theorem 3} (Busch, 2003) \textit{Any generalized probability measure $\mu$ is of the form $\mu(F)=\mathrm{trace}(\rho F)$ for
some density operator $\rho$.}
\bigskip

It is now easy to see that $q(F|\tau)$ on the likelihood effects of the previous Section is a generalized probability measure if Assumption 1 holds:
(1) follows since $q$ is a probability; (2) since $F=I$ implies that the likelihood is 1 for all values of the theoretical variable; finally (3) is a
consequence of the corollary of Theorem 2. Hence there is a density operator $\rho =\rho(\tau)$ such that
$q(F|\tau)=\mathrm{trace}(\rho(\tau)F)$ for all likelihood effects $F=F(\mathbf{u}; z ,\tau)$. This is a result that is valid for all experiments, and it can be seen as a first general version of Born's formula.

The problem of defining a generalized probability on the set of effects is also discussed in Busch et al. (2016).

Define now a \textit{perfect experiment} as one where the measurement uncertainty can be disregarded. The quantum mechanical literature operates
very much with perfect experiments which result in well-defined states $|j\rangle$.  From the point of view of statistics, if, say the 99\% confidence or
credibility region of $\theta^b$ is the single point $u_{j}^b$, we can infer approximately that a perfect experiment has given the result $\theta^b =u_{j}^b$.

In our epistemic setting then: We have asked the question: `What is the value of the accessible e-variable $\theta^b$?', and are
interested in finding the probability of the answer $\theta^b =u_{j}^b$ though a perfect experiment. If $u_j^b$ is a non-degenerate eigenvalue of the operator corresponding to $\theta^b$, this is the probability of a well-defined state $|b;j\rangle$.
Assume now that this probability is sought in a setting defined as follows: We have previous knowledge of the answer
$\theta^a =u_{k}^a$ of another maximal question: `What is the value of $\theta^a$?' That is, we know the state $|a;k\rangle$. ($u_k^a$ is non-degenerate.)

These two experiments, the one leading to $|a;k\rangle$ and the one leading to $|b;j\rangle$, are assumed to be performed in equivalent contexts $\tau$.
\bigskip

\textbf{Theorem 4} [Born's formula, simple version] \textit{ Assume a rational epistemic setting. In the above situation we have:}
\begin{equation}
P(\theta^b =u_{j}^b |\theta^a =u_{k}^a)=|\langle a;k|b;j\rangle|^2 .
\label{Born}
\end{equation}

An advantage of using the version of Gleason's theorem due to Busch (Theorem 3) in the derivation of Born's formula, is that this version is valid also in dimension 2. Other derivations using the same point of departure, are Caves et al. (2004) and Wright and Weigert (2019). In Wright and Weigert (2021) the class of general probabilistic theories which also admit Gleason-type theorems is identified. But for instance, Auffeves and Granger (2019) derive the Born formula from other postulates.

\subsection{A survey of quantum probabilities}

Here are three easy consequences of Born's formula:
\begin{enumerate}
\item[1)] If the context of the system is given by the state $|a;k\rangle$, and $A^b$ is the operator corresponding to the maximal accessible variable $\theta^b$, then the expected value of a
perfect measurement of $\theta^b$ is $\langle a;k|A^b |a;k\rangle$.

\item[2)] If the context is given by a density operator $\rho$, and $A$ is the operator corresponding to the maximal accessible variable $\theta$, then the expected value of a
perfect measurement of $\theta$ is $\mathrm{trace}(\rho A)$. 

\item[3)] In the same situation the expected value of a perfect measurement of $f(\theta)$ is $\mathrm{trace}(\rho f(A))$.
\end{enumerate}

These results give an extended interpretation of the operator $A=A^\theta$: There is a simple formula for all expectations in terms
of the operator. On the other hand, the set of such expectations determine the state of the system. Also on the other hand: If $A$ is specialized to the operator of an indicator
function, we get back Born's formula, so the consequences are equivalent to this formula.

A consequence of 3) above is that $\theta=\theta^b$ does not need to be maximal in order that a Born formula should be valid. The version 2) of Born's formula is also valid when the basic variable $\theta^a$ behind the densitity operator $\rho$ is not maximal, under a certain assumption: Assume that $\theta^a=f(\lambda^a)$, where $\lambda^a$ is maximal and the conditional probability distribution of $\lambda^a$, given $\theta^a$ is uniform.

Measurements of empirical variables are discussed in Helland (2021), but here we will look at the simpler case of a perfect measurement. Assume that we know the state $|\psi\rangle$ of a system, and that we want to measure a new theoretical variable $\theta^b$. This can be discussed by means of the projection operators $ \Pi_j^b =|b;j\rangle\langle b;j|$. First observe that by a simple calculation from Born's formula
\begin{equation}
P(\theta^b=u_j^b |\psi)= \| \Pi_j^b |\psi\rangle \|^2.
\label{Born21}
\end{equation}
This formula is also valid when  $\Pi_j^b$ is a multidimensional projection.

It is interesting that Shrapnel et al. (2017)  recently simultaneously derived \emph{both} the Born rule and the well-known collapse rule from a knowledge-based perspective. The collapse rule is further discussed in Helland (2021), but in this article we will just assume this derivation as given.
Then, after a perfect measurement $\theta^b=u_j^b$ has been obtained, the state changes to
\[|b;j\rangle=\frac{\Pi_j^b |\psi\rangle}{\|\Pi_j^b |\psi\rangle\|}.\]

Successive measurements are often of interest. We find
\[
P(\theta^b=u_j^b\ \mathrm{and\ then}\ \theta^c=u_i^c|\psi)=P(\theta^c=u_i^c |\theta^b=u_j^b)P(\theta^b=u_j^b |\psi)\]
\[=\|\Pi_i^c \frac{\Pi_j^b |\psi\rangle}{\|\Pi_j^b |\psi\rangle\|}\|^2 \|\Pi_j^b |\psi\rangle \|^2= \|\Pi_i^c \Pi_j^b |\psi\rangle \|^2.\]

In the case with multiple eigenvalues, the formulae above are still valid, but the projectors above must be replaced by projectors upon eigenspaces. 

\subsection{Born's rule in a decision setting}

Go back to the person $C$ who is facing at least two different maximal decisions, with corresponding decision variables $\theta$ and $\eta$. Each of these decisions may be composed of several partial decisions; then the decision variables may be thought of as vectors. However, we will make the simplifying assumption that each partial decision has a finite number of outcomes, say possible actions; then the total decision also has a finite number of outcomes, and without loss of generality, we can order these actions as $\{a_i; i=1,2,..,r\}$.

Acording to Kahnemam (2011), some of our decisions are slow, and sometimes, in fact in very many cases, we have to make fast decisions, relying on earlier experience, perhaps in a subconscious way, relying on some abstract ideals that we may have in our minds at the moment of decision, or, alternatively, ideals that have determined our earlier decisions. This is consistent with the model assumptions that we have made above in connection to Born's rule. And the fact that such fast decisions may be modelled by quantum probabilities, is consistent with a lot of empirical findings, see Pothos and Busemeyer (2022).

Focus again on the person $C$ and his fast decisions. Depending upon the situation that he is in, he may just have done several irrelevant decisions, collected in the variable $\eta$, but at the same moment, he is trying to focus on the problem that he really is interested in, and the corresponding (maximal) variable is called $\theta$. There is not much lack of generality in assuming that these two decision processes have the same number of actions $r$; we can only add some abstract irrelevant actions to one of the decision variables.

According to my interpretation of quantum theory and of decision making, each of the two decision variables has an operator connected to it, and the eigenspaces of such an operator correspond to a concrete outcome of the decision process. The probabilities are calculated as in (\ref{Born21}). The application of these theoretical considerations is best studied by a simple example.

\section{Quantum decision theory in practice; an example}

Consider a situation with one doctor and one patient. The doctor has the choice between two mutually excluding medicines $A$ and $B$.
\bigskip

The doctor starts by asking a number of question, and he obtains answers. Result: A state $|\psi\rangle$ for the patient. This is the starting point of my model of what is going on in the mind of the doctor during the decision process. The state is modelled by some vector in an underlying Hilbert space, connected to the mind of the doctor.
\bigskip

The doctor is interested in several epistemic probabilities:

$P(A\ \mathrm{helps})$, $P(B\ \mathrm{helps})$, 

$P(A\ \mathrm{helps\ and}\ B\mathrm{\ helps\ given\ that\ he\ knows\ that\ }A\mathrm{\ helps})$.

The simplest case is where these two medicines work completely independently:

\[P(A\ \mathrm{helps\ and\ also}\ B\mathrm{\ helps})=P(B\ \mathrm{helps\ and\ also\ }A\mathrm{\ helps})\]
\[=P(A\ \mathrm{helps})P(B\ \mathrm{helps}).\]

This can also in principle be modelled in quantum language by using the tensor product $\mathcal{H}^A\otimes\mathcal{H}^B$ with two obvious Hilbert spaces $\mathcal{H}^A$ and $\mathcal{H}^B$. But such a model will in this situation be unnecessarily complicated.
\bigskip

However, from our point of view, the most interesting case is where there is some coupling between the two medicines. The doctor must make fast decisions. In his mind he may have a lot of other decisions that he has made or is going to make. As a model, these may be collected in a decision variable $\eta$. But now he focuses on the actual patient. The decisions that he is going to make on this patient, are modelled, and collected in the decision variable $\theta$. In the doctor's mind at the moment in question, is also all the information that he posesses about the medicines $A$ and $B$. As a doctor, he has ideals learned through many years of education and experience; these ideals can be modelled in terms of some abstract perfectly rational being $D$.

In agreement with Theorem 1, I will propose a quantum model with 
Hilbert space $\mathcal{H}$. The projector upon the subspace indicating that $A$ helps, is denoted by $\Pi_A$.
\smallskip

From Born's rule we get:

\[P(A\ \mathrm{helps})=\|\Pi_A|\psi\rangle \|^2 =\langle\psi | \Pi_A |\psi\rangle.\]
\smallskip

The new state after the doctor knows that $A$ helps is now:

\[|\psi_A\rangle = \frac{\Pi_A |\psi\rangle}{\|\Pi_A |\psi \rangle \|}.\]

To continue the model, consider two orthogonal spaces in $\mathcal{H}$:
$V_B$: Indicating that $B$ helps. The projector upon this space is $\Pi_B$.
And $V_B^{\mathrm{perp}}$: Indicating that $B$ does not help. The projector upon this space is $\Gamma_B = I-\Pi_B$.
\bigskip

This gives
\[P(A\ \mathrm{helps\ and\ also}\ B\mathrm{\ helps})=\]
\[P(A\ \mathrm{helps})P(B\ \mathrm{helps}|A\ \mathrm{helps})=\]
\[\|\Pi_A\psi\|^2 \| \Pi_B \psi_A \|^2 = \|\Pi_B \Pi_A |\psi\rangle\|^2.\]

\subsection{Some consequences}

The marginal probability that $B$ helps is $P(B\ \mathrm{helps})=\|\Pi_B|\psi\rangle \|^2$. Then by a geometric argument it can be seen that there exist situations where
\[P(A\ \mathrm{helps\ and\ also}\ B\mathrm{\ helps})>P(B\ \mathrm{helps}),\mathrm{\ i.e.,}\] 
\[\|\Pi_B \Pi_A|\psi\rangle \| > \|\Pi_B|\psi\rangle\|.\]
For a tentatively illustration, see Figure 1.

\begin{figure}
\includegraphics[width=0.5\textwidth]{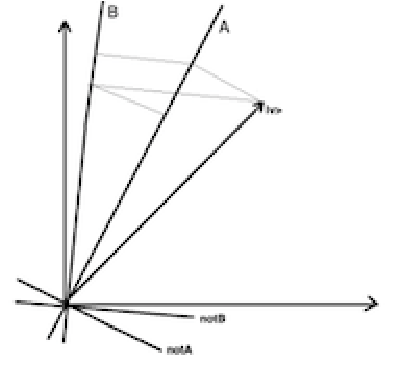}
{\caption{Geometry of two spaces.\label{fig1}}}
\end{figure}

Also, the law of total probability does not hold. There is additivity in the probability amplitudes:
\[\Pi_A |\psi\rangle =\Pi_B \Pi_A |\psi\rangle +\Gamma_B \Pi_A |\psi\rangle =a+b,\]
so that
\[P(A\ \mathrm{helps})=\|\Pi_A|\psi\rangle \|^2  =|a|^2 +|b|^2 +a^*b+ab^*.\]
But
\[P(A\ \mathrm{helps\ and\ also}\ B\mathrm{\ helps}) + P(A\ \mathrm{helps\ but}\ B\mathrm{\ does\ not\ help})\]
\[=|a|^2+|b|^2.\]

As a special case of the violation of the law of total probability: The sure thing principle does not hold in quantum decision theory.

Finally, note that one my have

\[P(A\ \mathrm{helps\ and\ also}\ B\mathrm{\ helps})\ne P(B\ \mathrm{helps\ and\ also\ }A\mathrm{\ helps}).\]

In general, events do not necessarily commute in relation to calculation of quantum probabilities.

All these results are well known properties of quantum probabilities; these probabilities simply do not follow Kolmogorov's laws.They are also well known properties in various empirical examples given in the literature of Quantum Decision Theory. My task now will be to relate the quantum probabilities, as observed in practice, to the assumptions made earlier in this article.

\section{Discussion}

\subsection{A summary of assumptions}

It is crucial that this article relies upon my earlier work on the foundation of quantum theory. This work in turn relies on several assumptions, as stated here in connection to the discussion around Theorem 1 in Chapter 2: 

1) The variables, here decision variables, are divided into accessible and inaccessible ones. A decision variable is said to be accessible to a person $C$ if and only if $C$ is able to carry out the corresponding decision. 

2) All accessible decision variables can be seen as functions of some large inaccessible theoretical variable. In practice this may be taken to mean that all earlier experiences of $C$, among other things related to his earlier decisions, together with what may be called his free will, are collected as a part of an hypothetical large variable $\phi$. Note that $C$ has not grown up in isolation. He is the result of interaction with other people, in particular the result of a specific human culture. Depending upon circumstances, he may have made specific conscious decisions to free himself from this culture, but nevertheless, the background is there. 

3) There exist maximal accessible decision variables. As stated earlier, these variables are connected to really difficult decisions. Such decisions may be conscious or unconscious. I will see it as fruitful to distinguish between what Kahneman (2011) calls decisions made by fast and slow thinking. Many of our decisions have to be fast, and I see the theory of this article as specially connected to these fast decisions.

4) To lead to a specific quantum theory, we may look at situations where there are two different, mutually exclusive decision processes in the mind of $C$ such that both can be seen as maximal. He or she then has to make a choice: Focus on the decision process connected to the decision variable $\theta$ or focus on the decision process connected to the decision variable $\eta$. Again, this choice may be conscious or unconscious. I will claim that at least in some cases, it is a fast decision, coupled solely to the unconsciousness of $C$.
\bigskip

In addition to these 4 assumptions, this article also relies on specific assumptions behind the Born rule, the means for calculating quantum probabilities -see Chapter 2 and Chapter 3:

5) The person with whom the probabilities are associated, believes in the statistical likelihood principle. Most statisticians believe in this principle, it is connected to the socalled Birnbaum's theorem: It can be deduced from simpler, more inuitive principles, see Helland (2021).

6) Both $C$ and the person with whom the probabilities are associated have ideals, ideals that can be modelled by some abstract being $D$, which is seen by the actors to be perfectly rational, as made explicit by the Dutch Book Principle. For different persons, these ideals may be very different. They may be connected to other, concrete persons, or they may just be abstract. This point is very crucial for me. We all go through life having certain ideals, and in most cases, this has a very positive effect on our lives.

But the ideals may also be connected to persons that I myself have little sympathy for. Extreme cases exist, like Adolf Hitler in Germany before and under the second world war, or Mafia bosses behind criminal decisions made by gang members. Extending this to my own opinions, I am also tempted to point at Vlademir Putin behind certain decisions made by 70 percent of the Russian population, and Donald Trump behind many of the decisions made by rightwing Americans.

What are then my own ideals? During my work with the quantum foundation, I have had a strong religious faith, as discussed explicitly earlier in this book and in the two articles Helland (2022d) and Helland (2023d). I do believe in the existence of a God, and I have sympathies with the best sides of all religious beliefs. Personally, I have chosen to be a Christain, but I do well understand people that have made other choices here. Religion is definitely connected to culture.

In making concrete decisions, I also rely on various ideals related to specific persons that I look up to.

\subsection{Decisions, learning, and cultures}

As stated in the Introduction, there is a close connection between decisions and learning. The learning processes start when we are infants, and continues throughout our lifes. The decisions that we make in our daily life can be seen as a product of this.

That implies that the decisions that we make today, partly may be a product of the interactions that we thoughout life have made with other people. In a somewhat larger perspective, the sum of all this may be called culture. The author Ralph B Stacey has written in that connection: `Culture is a set of attitudes, opinions and convictions that a group of people share, on how one shall behave, how things shall be evaluated and be done, what questions are important and what answers that can be accepted. The most important elements in a culture are unconscious, and cannot be changed from the outside.'

It should be emphasized that belonging to a culture has many positive sides, but a culture may become highly negative if it leads to hate and mistrust of people from other cultures.

It should be mentioned in this connection that Helland (2024b) is an attempt to create common results based on three different scientific cultures. I see science as a very important background for several of our decisions, but I see it as negative that, partly, science may be divided into disjoint cultures, with too few attempts of explicit communication between the different cultures.

\subsection{Everyday decisions}

Most everyday decisions are not related to a maximal decision variable. But sometimes a decision requires the most of our resources, say in certain situations during driving a car. Note that my discussion of Quantum Decision Theory, in particular the discussion of Born's rule, is not only valid when the variables are maximal. This is an important point, and it is perhaps not properly emphasized in the discussion above. Again I refer to Mogiliansky et al. (2009) for theoretical and practical implications.

This should imply that Quantum Decision Theory applies to many everyday decisions, and this is confirmed empirically by the QDT literature.

What Born's rule requires, is some observer to whom the probabilities are connected, a person that that has a mental model that determines these probabilities. In the QDT literature, this is represented by the researcher that carries out the relevant investigation.

The exception to this use of Born's rule in a decision context, may be what Kahneman (2011) calls decisions based upon slow thinking, decision where the actor $C$ actively uses his conscious self in the process of making the decisions. It seems unlikely that such decisions can be simply modeled by QDT.

\subsection{Decisions made by political leaders}

Political leaders have to make decisions that may have consequences for many people. This gives them a great responsibility. Most democratic leaders are able to live up to this responsibility, but there are exceptions.

These responsibilities are spread across a large number of concrete issues. One important issue is the climate question. It is a hope that political leaders in all countries now may be able to make the right decisions here.

Many of the issues that political leaders have to make decision on, are really difficult, often too difficult to be left to a single person. In democratic countries, joint decisions are often made by a group of communication persons. Such joint discussions are not discussed in the present chapter, but they could have been, in the light of a version of Theorem 1 which is valid for a group of communicating actors, see Helland (2024a).

It is a hope that the really important decisions made by political leaders are based on Kahneman's slow thinking, conscious decisions that are outside the scope of this article.

Certain political leaders have made decisions that really have cruel consequences. Some examples are 

1) The cruel attack on Ukraine initiated by Vlademir Putin in February 2022. The war is continuing here, and through propaganda and other means, many Russians support the war. Putin and his followers seem to base much of their motivation on the notion of power and on mistaken judgements. Vladimar Putin claims to be a Christian, but he seems to have forgotten Christ's main command, urging us to love all humans in addition to loving God.

2) The attack on innocent sivilians ordered by the leaders of Hamas in October 2023; 

3) The continuous attacks on innocent people in Gaza initiated by Israel's leaders. This has escalated now to a terrible humanitarian situation. Benjamin Netanyahu and his rightwing gouvernment are the main responsible here, but unfortunately, in some of their decisions they are supported by a substancial part of Israel's population. 

4) The decisions taken by generals on the both sides in the confict in Sudan. These decisions have resulted in a quite unnecessary civilian war, which has implied terrible suffering for a large part of Sudan's population.

It is a hope that these leaders in due time will repent and realize that we all, in our final phase, may be subject to a divine  judgement.

One constant threat to humanity are the nuclear weapons, weapons that now are spread across too many countries. Any use of nuclear weapons will lead to enormous sufferings, and may in the worst case be the start of a process that can lead to the extinction of the human race. 

One clear wish is that they never should be used again. How can that be avoided? My idea here is that this may be connected to the observation that all people, also political leaders may have some religious feelings, at least some thoughts around their destiny after death. May be the following thought could be spread and made common: The political leader who ever dares to order some atomic bomb to be released over humans, deserves an extremely hard punishment after death. If this could be believed by all our leaders, the atomic arsenal spread around the world will be totally useless, and negotiations around the reduction of this arsenal might start. This is of course an idealistic thought, but on this important issue, one should really try to think in a new way.

\section{Concluding remarks}

Quantum decision theory seems to be applicable in many cases where decisions are to be made, primarily in the cases which Kahneman (2011) calls decisions made through fast thinking, thinking of type 1.  When a prior is available, an alternative is a Bayesian decision theory. This theory will follow the Kolmogorov rules, and apart from the existence of a prior, the theory does not require any special assumptions.

In this article, I have concentrated on decisions made by a single person. But the theory can be extended to joint decisions made by a group of communicating persons. Then these persons must have common ideals, as modelled by a perfectly rational higher being $D$. The ideals must be common in the setting determined by the relevant decision variables.

It is important that the rules of quantum probabilities are different from the usual Kolmogorov probability rules. The law of total probability does not hold, and the probabilities related to two events may depend on the order in time of the events.

As a basis for quantum decision theory as discussed here, lies the idea of two complementary decision variables at the same time in the mind of some person or in the joint minds of a group of communicating persons. I repeat that two variables are here called complementary if they are really different (not in one-to-one correspondence), have similar ranges, and are maximal as accessible variables. Another essential assumption is the existence of ideals that are perceived as perfectly rational, modelled by a hypothetical `ideal higher being'.

The quantum probabilities may be though of as relative to this `ideal higher being', hence they must be looked upon as absolute probabilities, not conditioned in any other way than by our past knowledge. This is confirmed by empirical findings in the quantum decision literature.

There are cases where powerful political leaders may be roughly modelled as such `ideal higher beings', relative to the influence they have on other people and their decisions. A fundamental requirement should then be that the leaders obey certain basic humanitarian conditions, including respect for other people's life. Unfortunately, this goal is not always satisfied in practice.

Examples where the assumptions behind decisions lead to conflicts, may be found within the mind of any person, but also between persons and between groups of persons.

These assumptions may in certain situations  be satisfied by very many different pairs of decision variables, perhaps leading to mental difficulties in the first case and to partly complementary world views in the last two cases. Then in these cases it might imply deep open conflicts between those persons or groups if some necessary conditions are there. We can see many examples of this phenomenon in the world today.

It is a hope that ideas related to this chapter may lead to an increased focus on making the best decisions for all people, for those given by political leaders in a global perspective, the best decisions in order to obtain peace on this earth and a viable future for all human beings.

\section*{Aknowledgements}

I first want to thank Solve S\ae b\o\ and Trygve Alm\o y for discussions. This article is partly based upon Chapter 11 in Helland (2025c). I am grateful for permission given by the editors here.

\section*{Conflicts of interest statement}

There are no conflicts of interest.

\section*{Data availability statement}

There are no data in this article.

\section*{References}

Aerts, D. (2009). Quantum structure in cognition \textit{Journal of Mathematical Psychology} \textbf{53}, 314-348.

  Allais, M. (1953). La psychologie de l'homme rationnel devant le risque: Critique des postulats et axiomes de l'ecole Ame\'{e}ricaine. \textit{Econometrica} \textbf{21}, 515-550.

 Ashtiani, M. and Azgomi, M.A. (2015). A survey of quantum-like approaches to decision making and cognition. \textit{Mathematical Social Sciences} \textbf{75}, 49-80.

 Auffeves, A. and Granger, P. (2019). Deriving Born's rule from an inference to the best explanation. arXiv:1910.13738 [quanr-ph].

 Busemeyer, J.R. and Bruza (2012). \textit{Quantum Models of Cognition and Decision.} Cambridge University Press, Cambridge.

 Busch, P. (2003). Quantum states and generalized observables: A simple proof of Gleason's Theorem.  \textit{ Physical Review Letters, 91}(12), 120403.

 Busch, P., Lahti, P., Pellonp\"{a}\"{a}, J.-P., \& Ylinen, K. (2016). \textit{Quantum measurement}. Springer, Berlin

 Campanella, M., Jou, D., \& Mongiovi, M. S. (2020). \textit{Interpretative Aspects of Quantum Mechanics.} Matteo Campella's Mathematical Studies. Cham, Switzerland: Springer.

 Caves, C.M., Fuchs, C.A., Manne, K., and Renes, J.M. (2004). Gleason-type derivations of the quantum probability rule for generalized measurements. \textit{Foundations of Physics} \textbf{34}, 193-209.

 Ellsberg, D. (1961). Risk, ambiguity and the Savage axioms. \textit{Quaterly Journal of Economics}. \textbf{75}, 643-679.

  Friston, K.J., Daunizeau, J. and Kiebel, S.J. (2009). Reinforcement learning or active inference? \textit{PLoS ONE} \textbf{4} (7) e6421.
  
  Haven, E. and Khrennikov, A. (2013). \textit{Quantum Social Science.} Cambridge University Press, Cambridge.

 Helland, I.S. (2021). \textit{Epistemic Processes. A Basis for Statistics and Quantum Theory.} Springer, Cham, Switzerland.

 Helland, I.S. (2022a). On the diversity and similarity of mathematical models in science. \textit{American Review of Mathematics and Statistics} \textbf{10} (1), 1-10.

 Helland, I.S. (2022b). On reconstructing quantum theory from two related maximal conceptual variables. \textit{International Journal of Theoretical Physics} \textbf{61}, 69. Correction (2023) \textbf{62}, 51.

 Helland, I.S. (2022c). The Bell experiment and the limitations of actors. \textit{Foundations of Physics} \textbf{52}, 55.
 
 Helland; I.S. (2022d). On religious faith, Christianity, and the foundation of quantum mechanics. \textit{European Journal of Theology and Philosophy} \textbf{2} (1), 10-17.

 Helland, I.S. (2023a). \textit{On the Foundation of Quantum Theory. The Relevant Articles.} Eliva Press, Chisinau, Moldova.

 Helland, I.S. (2023b). On the Bell experiment and quantum foundation. \textit{Journal of Modern and Applied Physics} \textbf{6} (2), 1-5.

 Helland, I.S. (2023c). Possible connections between relativity theory and a version of quantum theory based upon theoretical variables. arXiv: 2305.15435 [physics.hist-ph].
  
 Helland, I.S. (2023d). Quantum mechanics as a theory that is consistent with the existence of God. \textit{Dialogo Conferences and Journal} \textbf{10} (1), 127.134.
  
 Helland, I.S. (2024a). An alternative foundation of quantum mechanics. arXiv: 2305.06727 [quant-ph]. \textit{Foundations of Physics} \textbf{54}, 3.
 
 Helland, I.S. (2024b). On probabilities in quantum mechanics. arXiv: 2401.17717 [quant-ph]. \textit{APL Quantum} \textbf{1}, 036116.
 
 Helland, I.S. (2024c). A new approach towards quantum foundation and some consequences. arXiv: 2403.09224 [quant-ph]. \textit{Academia Quantum} \textbf{1}, 7282 .
 
Helland, I.S. (2025a). Some mathematical issues regarding a new approach towards quantum foundation. \textit{Journal of Mathematical Physics} \textbf{66}, 092103.

Helland, I.S. (2025b). Quantum probability for statisticians; some new ideas. \textit{Methodology and Computing in Applied Probability} \textbf{27} (84), 1-24.

Helland, I.S. (2025c). \textit{On God, Complementarity, and Decisions. Consequences of a New Approach towards Quantum Foundation.} Ethics International Press, UK.
  
 Helland, I.S. (2026). Towards optimal linear prediction. Discussion article. \textit{Scandinavian Journal of Statistics} \textbf{53} (1), 1-32.

 Kahneman, D. (2011). \textit{Thinking, Fast and Slow} Penguin Books. London.

 Kahneman, D., Sibony, O. and Sunstein, C.R. (2021). \textit {Noise. A Flaw in Human Judgement.} Brockman, Inc., New York.
 
 Khrennikov, A. (2010). \textit{Ubiquitous Quantum Structure} Springer, Berlin.
 
 Mogiliansky, A.L., Zamir, S. and Zwirn, H. (2009). Type indeterminacy: A model of the KT(Kahneman-Tverski)-man. \textit{Journal of Mathematical Psychology} \textbf{53} (5), 349-361.

 Newell, B.R., Lagnado, D.A. and Shanks, D.R. (2015). \textit{Straight Choices. The Psychology of Decision Making.} 2. edition. Psychology Press, London.

 Pothos, E.M. and Busemeyer, J.R. (2022). Quantum cognition. \textit{Annual Review of Psychology} \textbf{73}, 749-778.

 Shrapnel; S., Costa, F. \& Milburn, G. (2017). \textit{Updating the Born rule.} arXiv: 1702.01845v1 [quant-ph].

 von Neumann, J. (1927). Wahrscheinlichkeitstheoretischer Aufbau der Quantenmechanik.  \textit{Nachrichten von der Gesellschaft der Wissenschaften zu G\"{o}ttingen, Mathematisch-Physikalische Klasse, 1927}, 245--272.
 
Wang, Z., Busemeyer, J.R., Altmanspacer, H., and Pathos, E.M. (2013). The potential of using quantum theory to build models of cognition. \textit{Topics in Cognitive Sciences} \textbf{5}, 672-688.

 Wright, V.J. and Weigert, S. (2019). A Gleason-type theorem for qubits based on mixtures of projective measurements. \textit{ journal of Physics A: Mathematical and Theoretical} \textbf{52} (5), 055301.

 Wright, V.J. and Weigert, S. (2021). General probabilistic theories with a Gleason-type theorem. \textit{Quantum} \textbf{5}, 588.

  Yukalov, V.I. and Sornette, D. (2014). How brains make decisions. \textit{Springer Proceedings in Physics} \textbf{150}, 37-53.

 Zwirn, H. (2016). The measurement problem: Decoherence and convivial solipsism. \textit{Foundations of Physics} \textbf{46}, 635-667.

 Zwirn, H. (2020). Nonlocality versus modified realism.
\textit{Foundations of Physics} \textbf{50}, 1-26.

\end{document}